\documentclass[12pt]{article}
\usepackage{amsfonts}
\usepackage{amssymb}
\usepackage{color}
\usepackage{graphicx}

\newcommand{\A}{{\mathfrak A}}

\newcommand{\F}{{\cal F}}

\newcommand{\mc}{\mathcal}

\newcommand{\be}{\begin{equation}}
\newcommand{\en}{\end{equation}}
\newcommand{\bea}{\begin{eqnarray}}
\newcommand{\ena}{\end{eqnarray}}
\newcommand{\beano}{\begin{eqnarray*}}
\newcommand{\enano}{\end{eqnarray*}}

\newcommand{\Zh}{{\hat Z}}
\newcommand{\Xh}{{\hat X}}
\newcommand{\Fh}{{\hat F}}
\newcommand{\Eh}{{\hat E}}
\newcommand{\BB}{\mc B}
\newcommand{\EE}{\mc E}
\newcommand{\pt}{{\Psi(t)}}

\newcommand{\FF}{\mc F}

\newcommand{\1}{1 \!\! 1}

\newcommand{\Hil}{\mc H}

\newtheorem{thm}{Theorem}

\newtheorem{lemma}[thm]{Lemma}
\newtheorem{prop}[thm]{Proposition}
\newtheorem{defn}[thm]{Definition}

\catcode `\@=11 \@addtoreset{equation}{section}
\def\theequation{\arabic{section}.\arabic{equation}}
\catcode `\@=12
\begin{document}

\begin{center}
{\Large \textbf{A  dynamical approach to compatible and incompatible questions}} \vspace{2cm%
}\\[0pt]

{\large F. Bagarello}
\vspace{3mm}\\[0pt]
DEIM, Facolt\`{a} di Ingegneria,\\[0pt]
Universit\`{a} di Palermo, I - 90128 Palermo,  and\\
\, INFN, Sezione di Napoli, Italy.\\[0pt]
E-mail: fabio.bagarello@unipa.it\\[0pt]
home page: www1.unipa.it/fabio.bagarello \vspace{8mm}\\[0pt]
\end{center}

\vspace*{2cm}

\begin{abstract}
\noindent We propose a natural strategy to deal with compatible and incompatible binary questions, and with their time evolution. The strategy is based on the simplest, non-commutative, Hilbert space $\Hil=\mathbb C^2$, and on the (commuting or not) operators on it. As in ordinary Quantum Mechanics, the dynamics is driven by a suitable operator, the Hamiltonian of the system. We discuss a rather general situation, and analyse the resulting dynamics if the Hamiltonian is a simple Hermitian matrix.
\end{abstract}

\vfill

\newpage


\section{Introduction}

Decision Making is quite an exciting area of research; it includes aspects from mathematics, physics, biology, neuroscience, psychology, etc. Understanding how a brain produces a decision, which are the mechanisms, what is needed and what is not during the procedure of decision, requires a lot of work and many more (and more refined) attempts than those existing nowadays in the literature. We refer to \cite{kelly,korner,maaser,white} for a small list of contributions in this area, contributions of  different kind, with different techniques, and with different perspectives. Recently, an increasing group of people started to use a quantum-like approach to Decision Making, and they began to explore the consequences of this approach, see \cite{Asanobook,busebook,buse3,khre1,KHR,khre2} for instance, and references therein. For instance, an interesting application to order effects is discussed in \cite{paulina}.

In a recent paper the possibility of using commuting (or not) operators has been discussed in connection with two relevant problems in Decision Making: the analysis of compatible and incompatible questions and order effects, \cite{bbkp}. The key idea was to use two different deformation matrices to work in the Hilbert space $\hat\Hil=\Bbb C^4$ and to use a single parameter $\theta$ as a {\em measure} of the incompatibility of the questions and of the relevance of the order. These were measured, in a sense, by making use of the Heisenberg-Robinson inequality in $\hat{\Hil}$. This approach was successfully used to explain some experimental data. However, no time evolution was considered in \cite{bbkp}, also because it was not so clear how to introduce the dynamics in that settings\footnote{We will return on this aspect later on.}. Another evident weak aspect of what is proposed in \cite{bbkp} is the impossibility of fixing uniquely the analytic form of the deformations, whose expression is suggested by certain natural requirements, but on the other hand could be quite general.

Here we adopt a similar approach, based again on the Heisenberg-Robinson inequality, but we show that it is enough to stay in $\Hil=\Bbb C^2$, independently of the nature of the questions we are interested to consider: binary compatible or incompatible questions can  be both analyzed  in $\Hil$, and the dynamics can  be also introduced in the analysis of the system in a natural way. Also, we don't have to worry on how to fix the analytic expression of the deformation matrices mentioned above, simply because they are not needed.

The paper is organized as follows: in the next section we introduce the problem and we set up the mathematical framework relevant for its analysis, in absence of any time evolution. In particular, we discuss the differences and the similarities of this framework for questions which can be compatible or not. In Section \ref{sectdynamics} we propose a dynamics for the system, and we discuss a few consequences of this proposal. Section \ref{sectexample} contains an example, while our conclusions are given in Section \ref{sectconclusions}. To keep the paper self-contained and to understand better some of the tools used in the paper, we have added two Appendices: in the first one, we discuss few properties of variances while, in the second, we discuss the possibility to saturate the Heisenberg-Robinson inequality.

\section{Stating the problem at fixed time}\label{sect2}

Suppose Alice is asked two binary questions: {\bf $Q_1$}: are you happy? and {\bf $Q_2$}: do you have a job? The answers can only be "yes" or "no". They can be thought to be mutually related (Alice is happy because she has a job) or not (Alice is happy independently of having a job or not). In \cite{busepot} the authors used two different Hilbert spaces depending on the relation between  {\bf $Q_1$} and {\bf $Q_2$}. In \cite{bbkp} the authors showed that it is possible to work in a single Hilbert space, $\hat\Hil=\Bbb C^4$, independently of the relation between {\bf $Q_1$} and {\bf $Q_2$}. The price to pay was to introduce two different deformation matrices, rather non-unique, but constructed following natural requirements. In what follows we will show that we can do better than this, by restricting to a simpler Hilbert space, $\Hil=\Bbb C^2$, and we can avoid using these deformation matrices, while keeping the role of commutativity between operators representing the questions above unchanged, and essential in our analysis.

We begin introducing two Hermitian operators $\Fh$ and $\Eh$, $\Fh=\Fh^\dagger$ and $\Eh=\Eh^\dagger$, having both eigenvalues $\pm1$ and eigenvectors $\BB_\Fh=\{f_+,f_-\}$ and $\BB_\Eh=\{e_+,e_-\}$ respectively:
\be
\Fh f_\alpha=\alpha f_\alpha, \qquad \Eh e_\alpha=\alpha e_\alpha,
\label{21}
\en
$\alpha=\pm1$. Of course, $\left<f_\alpha,f_\beta\right> = \left<e_\alpha,e_\beta\right> =\delta_{\alpha,\beta}$. Moreover, since $\Fh$ and $\Eh$ have only two eigenvalues, it is natural to look at them as $2\times2$ matrices, acting on $\Hil$. Let $\Psi\in\Hil$ be a normalized vector, somehow describing Alice. We introduce the mean values of $\Fh$ and $\Eh$ on $\Psi$:
\be
\F_\Psi=\left<\Psi,\Fh \Psi\right>, \qquad \EE_\Psi=\left<\Psi,\Eh \Psi\right>,
\label{22}
\en
and we interpret these as Alice's {\em degree of happiness} and {\em degree of employment} in $\Psi$: the closer $\F_{\Psi}$ is to 1, the happier Alice is. If  $\F_{\Psi}\simeq0$, Alice is not happy at all! Analogously, if $\EE_{\Psi}\simeq1$, then Alice feels she is employed\footnote{She could have a part-time job, or a temporary employment.}. We also introduce the two related variances as follows:
\be
(\Delta \F_\Psi)^2=\left<\Psi,(\Fh-\F_\Psi)^2\Psi\right>=\|(\Fh-\F_\Psi)\Psi\|^2=\left<\Psi,\Fh^2 \Psi\right>-\F_\Psi^2,
\label{23}\en
and
\be
(\Delta \EE_\Psi)^2=\left<\Psi,(\Eh-\F_\Psi)^2\Psi\right>=\|(\Eh-\EE_\Psi)\Psi\|^2=\left<\Psi,\Eh^2 \Psi\right>-\EE_\Psi^2.
\label{24}\en
Following the ordinary interpretation in quantum mechanics, see also  Appendix A, we consider $\Delta \F_\Psi$ and $\Delta \EE_\Psi$ as the incertitude on $\F_\Psi$ and $\EE_\Psi$, respectively: the  smaller the values of $\Delta \F_\Psi$, the smaller the uncertainty on Alice's mood. This will be clarified by our results.

We first observe that, given $\Psi=c_+f_++c_-f_-$, $\|\Psi\|^2=|c_+|^2+|c_-|^2=1$, we get
\be
\F_\Psi=|c_+|^2-|c_-|^2=2|c_+|^2-1.
\label{25}\en
Also, recalling that $\BB_\Eh$ is an orthonormal (o.n.) basis, we can also write $\Psi=d_+e_++d_-e_-$, $\|\Psi\|^2=|d_+|^2+|d_-|^2=1$, and we get
\be
\EE_\Psi=|d_+|^2-|d_-|^2=2|d_+|^2-1.
\label{26}\en
These formulas imply that both $\F_\Psi$ and $\EE_\Psi$ belong to the closed interval $[-1,1]$:$$-1\leq  \F_\Psi,\EE_\Psi\leq1,$$
for all possible normalized $\Psi$. This is because, of course, the eigenvalues of $\Fh$ and $\Eh$ are $\pm1$. Otherwise, the range of variability of $\F_\Psi$ and $\EE_\Psi$ would be different.
The following result can now be proved:

\begin{prop}\label{prop1}
	The following statements are equivalent: ($F_1$) $\F_\Psi=\pm1$; ($F_2$) $\Psi=c_\pm f_\pm$, for some complex constants $c_\pm$ such that  $|c_\pm|=1$; ($F_3$) $\Delta \F_\Psi=0$.
	
	Similarly, the following statements are also equivalent: ($E_1$) $\EE_\Psi=\pm1$; ($E_2$) $\Psi=d_\pm e_\pm$, for some complex constants $d_\pm$ such that  $|d_\pm|=1$; ($E_3$) $\Delta \EE_\Psi=0$.
\end{prop}

{\bf Proof:--} 
Suppose first that $\F_\Psi=1$. Since $0\leq|c_\alpha|^2\leq1$, $\alpha=\pm1$, formula (\ref{25}) implies that $c_-=0$ and $|c_+|=1$, so that $(F_2)$ follows. Vice versa, if we have, for instance, $\Psi=c_+ f_+$, for some complex constant $c_+$ with  $|c_+|=1$, it is clear that $\Psi$ is a normalized eigenstate of $\Fh$, corresponding to eigenvalue +1. Formulas (\ref{21}) and (\ref{22}) easily imply that $\F_\Psi=1$. 

Suppose now that  $\Psi=c_+ f_+$, $|c_+|=1$. Hence, as we have seen, $\F_\Psi=1$, and we have $$(\Fh-\F_\Psi)\Psi=\Fh\Psi-\F_\Psi\Psi=\Psi-\Psi=0,$$ so that, see (\ref{23}), $\Delta \F_\Psi=0$. Viceversa, if $\Delta \F_\Psi=0$, then (\ref{23}) implies that $(\Fh-\F_\Psi)\Psi=0$, which means that $\Psi$ is an eigenstate of $\Fh$ with eigenvalue $\F_\Psi$, which can only be $\pm1$.

The other cases can be proved in a similar way.

\hfill$\Box$

The meaning of this proposition is the following: we can be sure of Alice's answer regarding $Q_1$ or $Q_2$ if, and only if, her {\em state of mind} $\Psi$ is an eigenstate of either $\Fh$, or $\Eh$, or both (which is possible only if $\Fh$ commutes with $\Eh$, see Section \ref{sectTROC}). When $\Psi$ is an eigenstate of, say, $\Fh$, the mean value of $\hat F$ on $\Psi$ is either +1 or -1, and, at the same time,  $\Fh\Psi=\pm\Psi$ and $\Delta F_\Psi=0$. On the other hand, if $\F_\Psi$ is neither +1 nor -1, then $\Psi$ is not an eigenstate of $\Fh$, and $\Delta F_\Psi>0$: our knowledge of Alice's status suffers of a double uncertainty: since  $\F_\Psi\in]-1,1[$,  we cannot really conclude that Alice is {\em totally} happy or not. Moreover, since $\Delta F_\Psi>0$, our knowledge is affected by an extra error related to this uncertainty, see Appendix A.

\subsection{The role of commutativity}\label{sectTROC}

So far we have considered $\Fh$ and $\Eh$ separately. What is relevant for us here is to connect the two operators and to see how our previous considerations can be extended when $\Fh$ and $\Eh$ are considered together. We begin with the following result.

\begin{prop}\label{prop2}
	The following results hold: (1) if $\F_\Psi=\EE_\Psi=\pm1$ then $f_\alpha=\gamma_\alpha e_\alpha$, $|\gamma_\alpha|=1$, $\alpha=\pm1$; (2) if $\F_\Psi=-\EE_\Psi=\pm1$ then $f_\alpha=\gamma_{\alpha,\beta} e_\beta$, $|\gamma_{\alpha,\beta}|=1$ if $\alpha\ne\beta$, and $\gamma_{\alpha,\alpha}=0$, $\alpha,\beta=\pm1$.
\end{prop}

{\bf Proof:--} 
Suppose that $\F_\Psi=1$. Hence, by Proposition \ref{prop1}, $\Psi=c_+ f_+$, with $|c_+|=1$. Moreover, since $\EE_\Psi=1$,  we  also have $\Psi=d_+ e_+$, with $|d_+|=1$. Hence $f_+=\frac{d_+}{c_+}\,e_+$, and $\left|\frac{d_+}{c_+}\right|=1$, as claimed. The other statements can be proved similarly.

\hfill$\Box$

This proposition can be {\em almost} inverted:

\begin{prop}\label{prop3}
	The following results hold: (1) if $f_\alpha=\gamma_\alpha e_\alpha$, $|\gamma_\alpha|=1$, $\alpha=\pm1$, then  $\F_\Psi=\EE_\Psi$; (2) if $f_\alpha=\gamma_{\alpha,\beta} e_\beta$, $|\gamma_{\alpha,\beta}|=1$ if $\alpha\ne\beta$, and $\gamma_{\alpha,\alpha}=0$, $\alpha,\beta=\pm1$, then $\F_\Psi=-\EE_\Psi$.
\end{prop}

{\bf Proof:--} 
Let $f_\alpha=\gamma_\alpha e_\alpha$, with $|\gamma_\alpha|=1$, $\alpha=\pm1$. Then the normalized $\Psi$ can be written as $\Psi=c_+f_++c_-f_-=c_+\gamma_{+}e_++c_-\gamma_{-}e_-$. Hence $\F_\Psi=|c_+|^2-|c_-|^2$, and $\EE_\Psi=|\gamma_{+}c_+|^2-|\gamma_{-}c_-|^2=|c_+|^2-|c_-|^2=\F_\Psi$.

The other statements can be proved similarly. 

\vspace{2mm}

\hfill$\Box$

{\bf Remarks:--} (1) It is clear that Proposition \ref{prop3} is not really the inverse of Proposition \ref{prop2}, since Proposition \ref{prop3} does not state that  $\F_\Psi$ and $\EE_\Psi$ only take values $\pm1$. This is not really surprising, since assuming that $f_\alpha=\gamma_\alpha e_\alpha$, $|\gamma_\alpha|=1$, does not imply anything on $\Psi$, so that $\F_\Psi$ can be easily different from $\pm1$. For instance, if we take $\Fh=\Eh$, $f_\alpha=e_\alpha$, and $\Psi=\frac{1}{\sqrt{2}}(f_+-f_-)$, we can check that $\F_\Psi=\EE_\Psi=0$.

(2) Proposition \ref{prop2} can be restated by saying that, if $|\F_\Psi|=|\EE_\Psi|=1$, then $\Fh$ and $\Eh$ have a common set of eigenvectors. Moreover, Proposition \ref{prop3} implies that, if $\Fh$ and $\Eh$ have a common set of eigenvectors, then $|\F_\Psi|=|\EE_\Psi|$, but they are not necessarily equal to 1. A consequence of this result is given by Proposition \ref{prop4} below.

\vspace{2mm}

\begin{prop}\label{prop4}
	The following results hold: (1) if $[\Fh,\Eh]=0$, then  $|\F_\Psi|=|\EE_\Psi|$, for all $\Psi\in\Hil$; (2) if there exists $\Psi\in\Hil$, $\|\Psi\|=1$, such that $|\F_\Psi|=|\EE_\Psi|=1$, then $[\Fh,\Eh]=0$.
\end{prop}

{\bf Proof:--} (1) If $[\Fh,\Eh]=0$, then  $\Fh$ and $\Eh$ have a common set of eigenvectors, and Proposition \ref{prop3} implies that $|\F_\Psi|=|\EE_\Psi|$, independently of $\Psi$.

(2) If $|\F_\Psi|=|\EE_\Psi|=1$ for a given normalized $\Psi$, Proposition \ref{prop2} implies that it is possible to diagonalize $\Fh$ and $\Eh$ together. Hence they commute.

\hfill$\Box$

{\bf Remarks:--} (1) Notice that it may happen that $[\Fh,\Eh]\neq0$, and $|\F_\Psi|=|\EE_\Psi|$. However, these cannot be equal to 1, since this would be in contrast with claim (2) of Proposition \ref{prop4}. A similar example can be easily constructed. Let us introduce
$$
\Psi=\frac{1}{\sqrt{2}}\left(
\begin{array}{c}
i \\
1 \\
\end{array}
\right), \qquad \Eh=\left(
\begin{array}{cc}
1 & 0 \\
0 & -1 \\
\end{array}
\right), \qquad \Fh=\left(
\begin{array}{cc}
\cos2\theta & -\sin2\theta \\
-\sin2\theta & -\cos2\theta \\
\end{array}
\right).
$$ 
The matrices $\Fh$ and $\Eh$ commute only if $\sin2\theta=0$. When this is not so, then $[\Fh,\Eh]\neq0$. Nevertheless, a straightforward computation shows that $\F_\Psi=\EE_\Psi=0$.

(2) In Remark (1) after Proposition \ref{prop3} we have given an example of two commuting operators, $[\Fh,\Eh]=0$, such that $\F_\Psi=\EE_\Psi\neq\pm1$, according to claim (1) of Proposition \ref{prop4}.

(3) Statement $(2)$ implies the following: if $\Psi\neq0$ is a common normalized eigenstate of $\Fh$ and $\Eh$, then $[\Fh,\Eh]=0$. This is because, in this case,  $|\F_\Psi|=|\EE_\Psi|=1$. Our claim follows.

Our previous results, together with those in \cite{bbkp}, suggest to introduce the following definition.

\begin{defn}
	Two questions $\bf Q_1$ and $\bf Q_2$ are compatible if the related operators $\Fh$ and $\Eh$ commute: $[\Fh,\Eh]=0$. They are incompatible if $[\Fh,\Eh]\neq0$.
\end{defn}

It is clear that this definition works also in the case of questions with more than just two answers. In this case, the main difference is in the dimension of the Hilbert space, which will be greater than two. This may have consequence also in the mathematics of the problem, even if most of the results simply pass trough. This extension will be considered in a future paper.

\subsubsection{Compatible questions}

Let us consider first what happens if  $\bf Q_1$ and $\bf Q_2$ are compatible. Hence $[\Fh,\Eh]=0$ and, using Proposition \ref{prop4}, $|\F_\Psi|=|\EE_\Psi|$, for all $\Psi\in\Hil$. It is interesting to observe that, because of the commutativity between $\Fh$ and $\Eh$, a vector $\Psi$ cannot be eigenstate of just one of these operators.

\begin{lemma}
	Let $[\Fh,\Eh]=0$ and $\Psi\in\Hil$, $\Psi\neq0$. Then only one of the following possibilities holds true: (1) $\Psi$ is an eigenstate of both $\Fh$ and $\Eh$; (2) $\Psi$ is  an eigenstate  of neither $\Fh$,  nor $\Eh$.
\end{lemma}

{\bf Proof:--} Suppose that $\Psi$ is an eigenstate of $\Fh$ with eigenvalue +1: $\Fh\Psi=\Psi$. Hence, since $[\Fh,\Eh]=0$,  
$$
\Fh(\Eh\Psi)=\Eh(\Fh\Psi)=\Eh\Psi,
$$
which shows that the non zero vector $\Eh\Psi$ is also an eigenstate of $\Fh$ with eigenvalue +1. But this eigenvalue has multiplicity one. Hence $\Eh\Psi$ must be proportional to $\Psi$: $\Eh\Psi=\alpha\Psi$, for some real $\alpha$. Incidentally, since the eigenvalues of $\Eh$ are only $\pm1$, $\alpha$ is either +1 or -1. The other case ($\Fh\Psi=-\Psi$) can be analysed analogously.

\hfill$\Box$

In view of this lemma, we consider separately the following cases: (i) $\Psi$ is eigenstate of both $\Fh$ and $\Eh$; (ii) $\Psi$ is an eigenstate of neither $\Fh$, nor $\Eh$. In both cases we work under the assumption that $[\Fh,\Eh]=0$.

Case (i) is rather simple: because of Proposition \ref{prop1}, $\Delta \F_\Psi=\Delta \EE_\Psi=0$ and $|\F_\Psi|=|\EE_\Psi|=1$. Hence questions ${\bf Q_1}$ and ${\bf Q_2}$ can be answered with no uncertainty, together.

Case (ii) is {\em richer}. First of all, since $\Psi$ is not an eigenstate of $\Fh$, $\Delta \F_\Psi>0$. Moreover, since $\Psi$ is not an eigenstate of $\Eh$ either, $\Delta \EE_\Psi>0$ as well. Proposition \ref{prop4} implies that $|\F_\Psi|=|\EE_\Psi|$, which cannot be equal to 1 since this is only possible if $\Psi$ is a joint eigenvector of $\Eh$ and $\Fh$. Hence we have a double uncertainty: one on the mean values of the observables, and an additional one on their variances. The Heisenberg-Robinson inequality for $\Fh$ and $\Eh$ on $\Psi$,
\be\Delta \F_\Psi\,\Delta \EE_\Psi\geq\frac{\left|\left<\Psi,[\Fh,\Eh]\Psi\right>\right|}{2},
\label{disHR}\en
is not particularly useful, in this case, since $[\Fh,\Eh]=0$. In this case, in fact, we get $\Delta \F_\Psi\,\Delta \EE_\Psi\geq0,$
which is obviously always true. We could refine this inequality by recalling that the Heisenberg-Robinson inequality  arises as a consequence of the following, more detailed, inequality:
\be
(\Delta X_\Psi)^2(\Delta Y_\Psi)^2\geq \left<\Psi,\hat W\Psi\right>^2+\frac{\left<\Psi,\hat Z\Psi\right>^2}{4},
\label{27}\en
for all $\hat X=\hat X^\dagger$, $\hat Y=\hat Y^\dagger$, and for all $\Psi\in\Hil$. Here $\hat Z=-i[\hat X,\hat Y]$ and $$\hat W=\frac{1}{2}\left\{\hat X-\left<\Psi,\hat X\Psi\right>,\hat Y-\left<\Psi,\hat Y\Psi\right>\right\},$$
where $\{\hat A,\hat B\}=\hat A\hat B+\hat B\hat A$ is the anticommutator between $\hat A$ and $\hat B$,
see \cite{bagbook} for instance. Hence, in this case, inequality (\ref{27}) can be refined as follows:
\be\label{add1}
\Delta \F_\Psi\,\Delta \EE_\Psi\geq \left|\left<(\Fh-\F_\Psi)\Psi,(\Eh-\EE_\Psi)\Psi\right>\right|,
\en
which could be used, if needed, to get a better lower bound on $\Delta \F_\Psi$ and $\Delta \EE_\Psi$.

\vspace{2mm}

{\bf Remark:--} It may be interesting to observe that, after some algebra, $\left<\Psi,\hat W\Psi\right>=\Re\left(\left<\Psi,\hat X\hat Y\Psi\right>\right)-\left<\Psi,\hat X\Psi\right>\left<\Psi,\hat Y\Psi\right>$, where $\Re(z)$ is the real part of $z$. Hence, $\left<\Psi,\hat W\Psi\right>=0$ if $\Psi$ is an eigenstate of either $\hat X$ or $\hat Y$ (or both). This is in agreement with the fact that, in these cases, the uncertainty relation is saturated.

\vspace{2mm}

\subsubsection{Incompatible questions}\label{sectIQ}

In this case $[\Fh,\Eh]\neq0$. It is not possible to have simultaneously $\Delta \F_\Psi=\Delta \EE_\Psi=0$ since, otherwise, $\Psi$ would be a common eigenvector of $\Fh$ and $\Eh$. But this, as we have shown, would imply that 
$[\Fh,\Eh]=0$, which is false. Hence we can have the following situations: (i) $\Delta \EE_\Psi=0$ and $\Delta \F_\Psi\neq0$; (ii) $\Delta \F_\Psi=0$ and $\Delta \EE_\Psi\neq0$; (iii) $\Delta \EE_\Psi\neq0$ and $\Delta \F_\Psi\neq0$.

Let us consider case (i): hence $\Psi$ is an eigenstate of $\Eh$ but not of $\Fh$. We have no uncertainty on ${\bf Q_2}$, but we cannot be sure on ${\bf Q_1}$. As for the Heisenberg-Robinson inequality, this is trivial $(0\geq0)$, since $\left<\Psi,[\Fh,\Eh]\Psi\right>=0$, even if $[\Fh,\Eh]\neq0$. We refer to Section \ref{sectexample} for some examples of this situation. We also recall that, in the literature, $\Eh$ and $\Fh$ are said to be {\em weakly commuting} on $\Psi$, \cite{ozawa}.

\vspace{2mm}

{\bf Remark:--} the correction $\left<(\Fh-\F_\Psi)\Psi,(\Eh-\EE_\Psi)\Psi\right>$ in (\ref{add1}) does not affect the result, since this term is also equal to zero, as one can easily check.

\vspace{2mm}

Case (ii) is completely analogous, exchanging the roles of $\Fh$ and $\Eh$.

As for case (iii), this is the most {\em noisy} situation: both ${\bf Q_1}$ and ${\bf Q_2}$ do not produce sharp answers, on such a vector. Moreover, we have no chance to minimize to zero at least one of the variances. The best we can do, according to 
$$\Delta \F_\Psi\,\Delta \EE_\Psi\geq\frac{\left|\left<\Psi,[\Fh,\Eh]\Psi\right>\right|}{2},$$
is to look for a state which saturates the inequality.  In an infinite-dimensional Hilbert space, and for $\Fh$ and $\Eh$ satisfying $[\Fh,\Eh]=i\1$, this is a property, for instance, of the so-called coherent states, \cite{gazeaubook}, which are usually interpreted as the {\em most classical} among all the quantum states of a system. We refer to Appendix B for the analysis of this problem in the present context.

\section{Dynamics}\label{sectdynamics}

So far we have described a static situation: Alice is asked a question, and she gives an answer, corresponding to her state of mind. The answer can be sharp or not, depending on her mood at the time when the question is asked. If we ask her two questions together, something more can happen, depending on whether the questions are compatible or not. But Alice's mood and the compatibility of the questions can easily depend on time: she can give an answer now, but later the answer could be completely different. Stated differently, time is an essential variable in the procedure of decision making. This has been discussed by several authors, in many papers, see \cite{khren1,khren2,buse2},  to cite just a few. Time evolution is what we will discuss here, in our particular context, and with a rather simplified point of view, as we will clarify later. This is important, to make the model exactly solvable.

Following the standard prescription in quantum mechanics, see also \cite{khren1,khren2,bagbook,bagbook2,buse2}, we assume that the state $\Psi$ of the system $\A$\footnote{The only agent of $\A$ is Alice, and we are interested only in her happiness and in her having a job.} evolves according to a Schr\"odinger like equation
\be
i\dot{\Psi}(t)=H\Psi(t),
\label{31}\en
where $H=H^\dagger$ is the {\em Hamiltonian} of the system. We will assume in this paper that Alice is a {\em closed system}, with no interaction with her environment. As widely discussed in \cite{baglovsto,bhk,bagbook}, the environment can provide an efficient tool in decision making, but it usually requires the use of an infinite-dimensional Hilbert space. Here, to keep the situation simple, we do not consider any reservoir. Hence $H$ is a simple $2\times 2$ Hermitian matrix. Also, to simplify further the situation, we take it time-independent. This means that the dynamics of the system {\em does not adjust itself} according to some external (or internal) information, as it happens, for instance, in \cite{BDGO,bagbook2}. Hence the solution of (\ref{31}) is simply $\Psi(t)=e^{-iHt}\Psi_0$, where $\Psi_0$ is the state of the system at $t=0$\footnote{Notice that for us time is just an independent variable which labels, in a continuous way, Alice's evolution.}. 

If $\Xh=\Xh^\dagger$ is a generic observable of $\A$, its mean value and its variance on $\pt$ can be defined as usual:
\be
X_{\Psi(t)}=\left<\pt,\Xh\pt\right>, \qquad (\Delta X_\pt)^2=\left<\pt,(\Xh-X_\pt)^2\pt\right>,
\label{32}\en
which can be rewritten as
\be
X_{\Psi(t)}=\left<\Psi_0,\Xh(t)\Psi_0\right>, \qquad (\Delta X_\pt)^2=X^2_{\Psi(t)}-(X_{\Psi(t)})^2,
\label{33}\en
where $\Xh(t)=e^{iHt}\Xh e^{-iHt}$. It is easy to check that, in our two-dimensional Hilbert space, $X_{\Psi(t)}$ and $\Delta X_\pt$ are periodic functions, see Section \ref{sectexample} for a concrete example. This implies, when $\Xh$ is replaced by $\Fh$ or $\Eh$, that Alice's mood always oscillates between two opposite values. This can be unpleasant, but it is deeply connected to the absence of the environment, see \cite{baglovsto,bhk}, which can be interpreted as the absence of any additional information reaching Alice. However, we will restrict ourselves to this simple case, in this paper, postponing an extension to the more general situation in a future paper.

For concreteness, we could fix the form of $H$ as follows $H=\omega_E\Eh+\omega_F\Fh+\lambda H_I$, where $H_0=\omega_E\Eh+\omega_F\Fh$ and  $H_I$ are respectively the {\em free} and the {\em interaction} Hamiltonians. The parameters $\omega_E$ and $\omega_F$ are positive quantities. The reason why we call $H_0$  the free Hamiltonian is because, if $\lambda=0$ and if $[\Eh,\Fh]=0$, then $\Eh(t)=\Eh$ and $\Fh(t)=\Fh$: the free Hamiltonian does not change the relevant observables of $\A$, at least for compatible questions: this is, in fact, what we expect in absence of interactions of any kind.

In this paper we are particularly interested in considering the role of the compatibility of questions in the analysis of $\A$. For this reason, from now on, we will discuss what happens if $H$ assumes its simplest form, $H=H_0$, but assuming further that  $[\Eh,\Fh]\neq0$, i.e. working with incompatible questions\footnote{Of course, $H_I$ can assume different expressions depending on what we are interested in. For instance, $H_I$ could be constructed using suitable combinations of the happiness and the employment operators. However, this is not what is interesting for us, here. In fact, with our choice $H=H_0$, we can understand more clearly the role of the commutator $[\hat E,\hat F]$ in the time evolution of $\A$. We should also mention other, but similar, choices of $H$, considered in \cite{broe}, again in connection with decision making but not directly with order effects.}. This produces interesting results, while for compatible questions it is easy to check that, if $H=H_0$,
$$
\F_{\Psi(t)}=\F_{\Psi_0}, \quad \EE_{\Psi(t)}=\EE_{\Psi_0}, \quad \Delta F_\pt=\Delta F_{\Psi_0}, \quad \Delta E_\pt=\Delta E_{\Psi_0},
$$
which means  that, if $[\Fh,\Eh]=0$, the time evolution does not affect at all the original values of these quantities. Of course, even in this simple case and with this choice of $H$, a non trivial time evolution can still be obtained  if $\lambda\neq0$, i.e. in presence of some interaction Hamiltonian. Needless to say, a different choice of $H$ could produce rather different conclusions, but in this paper we will restrict to the form of $H$ introduced above.

Suppose then that $[\Eh,\Fh]\neq0$. We can always assume we are working on the eigenstates of $\Fh$,  $\BB_\Fh=\{f_+,f_-\}$. With this choice, $\Fh$ is represented by the following matrix:
$$
\Fh=\left(
\begin{array}{cc}
1 & 0 \\
0 & -1 \\
\end{array}
\right).
$$
Since the eigenstates of $\Eh$ form a different (in general) o.n. basis of $\Hil$, $\BB_\Eh=\{e_+,e_-\}$, and since all o.n. bases are unitarily equivalent, we conclude that $e_\alpha=Uf_\alpha$, $\alpha=\pm1$, for some unitary operator $U$: $U^\dagger=U^{-1}$. It is known that the most general $2\times2$ unitary matrix $U$ can be written as
$$
U=\left(
\begin{array}{cc}
a & b \\
-\overline{b}\,e^{i\varphi} & \overline{a}\,e^{i\varphi} \\
\end{array}
\right), \qquad \Rightarrow \qquad U^{-1}=\left(
\begin{array}{cc}
\overline{a} & -b\,e^{-i\varphi} \\
\overline{b} & a\,e^{-i\varphi} \\
\end{array}
\right),
$$ 
where $|a|^2+|b|^2=1$, and $\varphi\in\mathbb{R}$. Since $\Eh$ has also eigenvalues $\pm1$, it is easy to check that its most general form is the following:
\be
\Eh=U\Fh U^{-1}=\left(
\begin{array}{cc}
|a|^2-|b|^2 & -2ab\,e^{-i\varphi} \\
-2\overline{a}\,\overline{b}\,e^{-i\varphi} & |b|^2-|a|^2 \\
\end{array}
\right).
\label{add2}\en
We stress that the operators $\Fh$ and $\Eh$ introduced here are the most general pair of operators satisfying our requirements. Straightforward computations show that
$$
\Zh=-i[\Fh,\Eh]=-4i\left(
\begin{array}{cc}
0 & -ab\,e^{-i\varphi} \\
\overline{a}\,\overline{b}\,e^{i\varphi} & 0 \\
\end{array}
\right),
$$
which is manifestly Hermitian. We see that $\Zh=0$ if $a=0$ or $b=0$, which implies that $\Eh$ is diagonal while the only non zero elements in $U$ are along the principal or the secondary diagonal. Simple computations show that
$$
[\Fh,\Eh]=i\Zh, \qquad [\Fh,[\Fh,\Eh]]=2i\Fh\Zh, \qquad [\Fh,[\Fh,[\Fh,\Eh]]]=4i\Fh^2\Zh,
$$
and so on. Moreover
$$
[\Fh,\Eh]=i\Zh, \qquad [\Eh,[\Fh,\Eh]]=2i\Eh\Zh, \qquad [\Eh,[\Eh,[\Fh,\Eh]]]=4i\Eh^2\Zh,
$$
and so on. Then
$$
\Fh(t)=e^{iHt}\Fh e^{-iHt}=\Fh+it[H,\Fh]+\frac{(it)^2}{2!}[H,[H,\Fh]]+\frac{(it)^3}{3!}[H,[H,[H,\Fh]]]+\ldots,
$$
and a similar formula can be written for $\Eh(t)$. We deduce
\be
\Fh(t)=\Fh+\rho(t,H)\omega_E\Zh, \qquad \Eh(t)=\Eh-\rho(t,H)\omega_F\Zh,
\label{34}\en
where
\be
\rho(t,H)=\1+\frac{1}{2!}(2itH)+\frac{1}{3!}(2itH)^2+\ldots=\frac{e^{2itH}-\1}{2i}\,H^{-1}.
\label{35}\en
The fact that $H^{-1}$ exists is guaranteed if $\omega_E\neq\omega_F$. Going back to $\F_{\Psi(t)}$ and $\EE_{\Psi(t)}$ we find
\be\label{36}
\F_{\Psi(t)}=\F_{\Psi_0}+\frac{\omega_E}{2i}\left<\Psi_0,\left(e^{2itH}-\1\right)H^{-1}\Zh\Psi_0\right>
\en
and
\be\label{37}
\EE_{\Psi(t)}=\EE_{\Psi_0}-\frac{\omega_F}{2i}\left<\Psi_0,\left(e^{2itH}-\1\right)H^{-1}\Zh\Psi_0\right>,
\en
which show how the mean value of the operators $\Fh$ and $\Eh$ evolve in time. As for the time dependence of the variances, since $\Fh^2=\Eh^2=\1$, and since $\Psi(t)$ is normalized, using formulas in (\ref{33}) we find that
\be(\Delta F_\pt)^2=1-(\F_{\Psi(t)})^2, \qquad (\Delta E_\pt)^2=1-(\EE_{\Psi(t)})^2,
\label{add2}\en
where $\F_{\Psi(t)}$ and $\EE_{\Psi(t)}$ are given in (\ref{36}) and (\ref{37}).

A simple (but not completely trivial) computation shows that $\F_{\Psi(t)}$ and $\EE_{\Psi(t)}$ are both real, for all possible choices of $\Psi_0$. In particular, if $\Psi_0$ is an eigenstate of $H$  with eigenvalue $\lambda_0$, $H\Psi_0=\lambda_0\Psi_0$, then it is possible to conclude that $\left<\Psi_0,\Zh\Psi_0\right>=0$. In fact, if from one side we have
$$
\F_{\Psi(t)}=\left<e^{-iHt}\Psi_0,\Fh e^{-iHt}\Psi_0\right>=\left<e^{-i\lambda_0t}\Psi_0,\Fh e^{-i\lambda_0t}\Psi_0\right>=\left<\Psi_0,\Fh \Psi_0\right>=\F_{\Psi_0},
$$
on the other side, using (\ref{36}), we obtain
$$
\F_{\Psi(t)}=\F_{\Psi_0}+\frac{\omega_E\left(e^{2it\lambda_0}-1\right)}{2i\lambda_0}\left<\Psi_0,\Zh\Psi_0\right>.
$$
Hence, this is possible only if $\left<\Psi_0,\Zh\Psi_0\right>=0$, as stated. Similar conclusions also follow from $\EE_{\Psi(t)}$.

Notice that, if $[\Fh,\Eh]=0$, then $\Zh=0$ and formulas (\ref{36}) and (\ref{37}) collapse into $\F_{\Psi(t)}=\F_{\Psi_0}$ and $\EE_{\Psi(t)}=\EE_{\Psi_0}$, as expected again because of our previous results.

We see that the mean values and the variances of $\Fh$ and $\Eh$ oscillate with time, between some minimum and some maximum values: there is no a priori equilibrium. This is not a big surprise, since we know that some equilibrium is reached, for instance, if the system (Alice) is coupled to some infinitely extended reservoir, see \cite{bhk,bagbook} for instance. In fact it is not possible, in any finite dimensional Hilbert space, to deduce some dynamics which, in the long run (i.e., without restricting $t$ to some particular interval), is not periodic or quasiperiodic. This result, see \cite{bagbook2}, does not depend on the particular choice of $H$ we use, as long as $H$ is Hermitian.

\vspace{2mm}

{\bf Remark:--} As we have already discussed before, in \cite{bbkp} we have adopted a different point of view to deal with ${\bf Q_1}$ and  ${\bf Q_2}$ working in a larger Hilbert space independently of their nature. The idea was to use two deformation maps to deform\footnote{Here we are adopting a slightly different notation with respect to \cite{bbkp}. In particular we call $\Fh$  the operator which was called $H$ in \cite{bbkp}, since $H$ is, here, the Hamiltonian of the system, as {\em universally} indicated in the literature.} $\Fh$ into $\Fh_\theta$ and $\Eh$ into $\Eh_\theta$. Depending on the value of the deformation parameter $\theta$, $\Fh_\theta$ can commute with $\Eh_\theta$, or not. But, if we want to take into account also the time evolution, it is not completely clear if we have to evolve $\Fh$ first, and then deform $\Fh(t)$ getting in this way $(\Fh(t))_\theta$, or if we have to deform $\Fh$ first, getting $\Fh_\theta$, and then  evolve $\Fh_\theta$ in time, getting $\Fh_\theta(t)$. This is because, in general,  $\Fh_\theta(t)\neq (\Fh(t))_\theta$. Similar problems occur for $\Eh$. Of course, this is not an issue in the present approach, which therefore, at least in this perspective, works better than the one in \cite{bbkp}.

\vspace{2mm}

Formulas (\ref{36}) and (\ref{37}) can be used to relate the values of $\F_{\Psi(t)}$ and $\EE_{\Psi(t)}$ directly. In fact, easy computations show that
\be\F_{\Psi(t)}=\F_{\Psi_0}+\frac{\omega_E}{\omega_F}\left(\EE_{\Psi_0}-\EE_{\Psi(t)}\right).
\label{38}\en
Of course, since $\|\Psi(t)\|=\|\Psi_0\|=1$, the Schwarz inequality implies that $$|\F_{\Psi(t)}|\leq \|\Fh\|=1, \qquad \EE_{\Psi(t)}|\leq \|\Eh\|=1,$$
for all $t\geq 0$. This means that formula (\ref{38}) imposes some constraints on the system. For instance, it is impossible to have $\F_{\Psi_0}=1$, $\EE_{\Psi_0}=0.8$ and $\EE_{\Psi(t)}=0.3$. This is because, being $\omega_E$ and $\omega_F$ positive quantities, the right-hand side of (\ref{38}) would be greater than 1, which is impossible.

\section{An application}\label{sectexample}

In this section we discuss an application in details, starting with the static case and then considering what happens when we consider also the time dependence.

\subsection{Before time evolves}

We begin our analysis by considering first the simplest situation, i.e. the case in which it is particularly simple to identify the state of the system. This is the case, for instance, if we imagine that  $\F_{\Psi_0}=1$ and $\EE_{\Psi_0}=0.8$. This choice simplifies the situation, since it is only compatible with $\Psi_0=c_+f_+$, with $|c_+|=1$\footnote{We fix $c_+=1$ in the following.}. Hence, if $\Eh$ takes the form
\be
\Eh=\left(
\begin{array}{cc}
\cos2\theta & \sin2\theta \\
\sin2\theta & -\cos2\theta \\
\end{array}
\right)
\label{ex1}\en
discussed below as a particular case of (\ref{add2}), see Appendix B, it follows that $\EE_{\Psi_0}=\left<f_+,\Eh f_+\right>=\cos2\theta=0.8$, which implies that $\sin2\theta=\pm\,0.6$. Hence $\Eh$ can only be one of the following two matrices:
$$
\Eh_1=\frac{1}{5}\left(
\begin{array}{cc}
	4 & 3 \\
    3 & -4 \\
\end{array}
\right), \qquad \Eh_2=\frac{1}{5}\left(
\begin{array}{cc}
4 & -3 \\
-3 & -4 \\
\end{array}
\right).
$$
For concreteness, we will use in the following the matrix $\Eh_1$.

Formula (\ref{38}) allows us to compute $\EE_{\Psi(T)}$, for some $T>0$, once we know $\F_{\Psi_0}$, $\EE_{\Psi_0}$ and $\F_{\Psi(T)}$. Indeed, from (\ref{38}) we get
$$
\EE_{\Psi(T)}=\EE_{\Psi_0}+\frac{\omega_F}{\omega_E}\left(\F_{\Psi(T)}-\F_{\Psi_0}\right)=0.8+0.2\,\frac{\omega_F}{\omega_E},
$$
if $\F_{\Psi(T)}=0.8$. Needless to say, this is only compatible with $\omega_F<\omega_E$. If this constraint is not satisfied, the numbers we are considering here are simply not allowed. Formula (\ref{38}), and its consequence above, shows that there is a strong relation between the mean values of $\Fh$ and of $\Eh$. This is not surprising, because of the fact that ${\bf Q_1}$ and ${\bf Q_2}$ are incompatible. This can be made explicit by computing the commutator $[\Fh,\Eh]$, which is different from zero. However, since $\Psi_0$ is an eigenstate of $\Fh$, we have $\left<\Psi_0,[\Fh,\Eh]\Psi_0\right>=0$: $\Fh$ and $\Eh$ are weakly commuting on $\Psi_0$, \cite{ozawa}. Moreover, for the same reason,  $\Delta \FF_{\Psi_0}=0$ and the inequality in (\ref{disHR}) is trivially satisfied.

\vspace{2mm}

It is maybe more interesting to consider a situation in which neither $\F_{\Psi_0}$ nor $\EE_{\Psi_0}$ are $\pm1$. In this case, the state of the system cannot be an eigenstate of $\Fh$ or of $\Eh$. Suppose, to be concrete, that the answers to ${\bf Q_1}$ and ${\bf Q_2}$ are, respectively, $0.6$ and $0.8$. This means that Alice is in a state $\Psi_{12}$ such that $\F_{\Psi_{12}}=0.6$ and  $\EE_{\Psi_{12}}=0.8$. The vector $\Psi_{12}$ can be (almost) fixed by means of these numbers, which are also sufficient to deduce the value of the angle $\theta$ in (\ref{ex1}). Calling $x$ and $y$ the components of $\Psi_{12}$, and restricting for simplicity to $x,y\geq0$,  $\F_{\Psi_{12}}=\left<\Psi_{12},\Fh\Psi_{12}\right>=0.6$, together with the normalization constraint $x^2+y^2=1$, produces $\Psi_{12}=\frac{1}{\sqrt{5}}\left(
\begin{array}{c}
2 \\
1 \\
\end{array}
\right)$. If we use this vector in $\EE_{\Psi_{12}}=\left<\Psi_{12},\Eh\Psi_{12}\right>=0.8$ we can fix (non uniquely!) the value of $\theta$. The simplest choice is $\theta=\frac{\pi}{4}$, which corresponds to the following expression for $\hat E$:
$
\Eh=\left(
\begin{array}{cc}
	0 & 1 \\
	1 & 0 \\
\end{array}
\right),
$
which do not commute with $\Fh$: $[\Fh,\Eh]=2\left(
\begin{array}{cc}
0 & 1 \\
-1 & 0 \\
\end{array}
\right)$. Hence ${\bf Q_1}$ and ${\bf Q_2}$ are incompatible. Nevertheless, it is easy to check that $\left<\Psi_{12},[\Fh,\Eh]\Psi_{12}\right>=0$, so that $\Fh$ and $\Eh$ are weakly commuting on $\Psi_{12}$. As for the variances, from (\ref{add2}) we get $\Delta \F_{\Psi_{12}}=0.8$ and $\Delta \EE_{\Psi_{12}}=0.6$, so that the Heisenberg-Robinson inequality is trivially satisfied.

\vspace{2mm}

It is interesting to see now what happens if we {\em reverse the order} of ${\bf Q_1}$ and ${\bf Q_2}$. As it is discussed in several experiments, see \cite{bbkp} and references therein, this corresponds usually to different answers to the same questions. We refer to the cited papers for a psychological discussion on the reason behind this order effect. Here we just assume that, with this reverse order, we get different answers. In particular, let us assume that the answers are now  $\F_{\Psi_{21}}=0.8$ and  $\EE_{\Psi_{21}}=0.6$, where $\Psi_{21}$ is the new vector, in general different from $\Psi_{12}$. In fact, with the same procedure as before, requiring that $\F_{\Psi_{21}}=\left<\Psi_{21},\Fh\Psi_{21}\right>=0.8$ and $\EE_{\Psi_{21}}=\left<\Psi_{21},\Eh\Psi_{21}\right>=0.6$ produces the vector $\Psi_{21}=\frac{1}{\sqrt{10}}\left(
\begin{array}{c}
3 \\
1 \\
\end{array}
\right)$ and the same matrix $\Eh$ as before. This is reasonable, since while we expect that the state of the system can change if we exchange the order of the questions, it is natural to imagine that this does not happen with the matrices representing the questions the expressions of the operators are invariant under ${\bf Q_1}\rightleftarrows{\bf Q_2}$, while the vectors are not. The variances can finally be easily computed.

\vspace{2mm}

{\bf Remark:--} It is worth pointing out that this is not always possible. In fact, if we consider the choice $\F_{\Psi_{a}}=\EE_{\Psi_{a}}=0.5$, repeating the same approach as above would produce a new vector $\Psi_a$, different from $\Psi_{12}$ and $\Psi_{21}$ above, and a new expression for $\Eh$, different from the previous one, $\left(
\begin{array}{cc}
	0 & 1 \\
	1 & 0 \\
\end{array}
\right)$. Hence we would have two pairs of values, $(\F_{\Psi},\EE_{\Psi})=(0.6,0.8)$  and $(\F_{\Psi},\EE_{\Psi})=(0.5,0.5)$, giving rise to different expressions of (at least one of) the observables of $\A$. This suggests to consider these choices as {\em incompatible}: if one pair can be found in a concrete experiment, the other cannot. We think that this aspect of our approach should be better understood.

\subsection{The time evolution}

With the forms of $\Fh$ and $\Eh$ deduced above, the Hamiltonian looks like $H=\left(
\begin{array}{cc}
\omega_F & \omega_E \\
\omega_E & -\omega_F \\
\end{array}
\right)$. Calling $\Omega^2=\omega_F^2+\omega_E^2$, it is possible to check that $H^{2k}=\Omega^{2k}\1$, and $H^{2k+1}=\Omega^{2k}\,H$, $k=0,1,2,3,\ldots$. These equalities, replaced in the expression for $\Fh(t)$ in (\ref{34}), imply that, for instance,
$$
\Fh(t)=\Fh+\frac{\omega_E}{2\Omega}\left(\sin(2t\Omega)+\frac{1}{i\Omega}(\cos(2t\Omega)-1)H\right)\hat Z,
$$
with a similar formula for $\Eh(t)$. Restricting ourselves to the case in which, at $t=0$, the system is in the state $\Psi_{12}$, we conclude that 
$$
\F_{\Psi_{12}(t)}=\F_{\Psi_{12}}+\frac{\omega_E}{5\Omega^2}\left(\cos(2t\Omega)-1\right)\left(3\omega_E-4\omega_F\right),
$$
with a similar formula for $\EE_{\Psi_{12}(t)}$. It is clear that this is a periodic function, with period $T=\frac{\pi}{\Omega}$. In particular, for $t_k=Tk$, $k=0,1,2,\ldots$, we have $\F_{\Psi_{12}(t_k)}=\F_{\Psi_{12}}$. 

It is interesting to observe that the ratio between $\omega_E$ and $\omega_F$ in the Hamiltonian has consequences on the time evolution of $\Fh$, $\Eh$, and on their mean values. In particular, if $\omega_E\gg\omega_F$, we find
$$
\F_{\Psi_{12}(t)}\simeq\F_{\Psi_{12}}+\frac{3}{5}\left(\cos(2t\Omega)-1\right),
$$
while, if $\omega_E\ll\omega_F$,
$$
\F_{\Psi_{12}(t)}\simeq\F_{\Psi_{12}}.
$$
This last result can be easily understood: if $\omega_F$ is much larger than $\omega_E$, the Hamiltonian $H$ can be approximated with $H\simeq\omega_F\Fh$, which commutes with $\Fh$. Hence, in this limit, $\Fh(t)\simeq \Fh$, and $\F_{\Phi(t)}\simeq\F_{\Phi}$ for each vector $\Phi\in\Hil$. For the same reason, if $\omega_E\gg\omega_F$, $H\simeq \omega_E \Eh$, and $\Eh(t)\simeq\Eh$. Hence $\EE_{\Phi(t)}\simeq\EE_{\Phi}$ for each vector $\Phi\in\Hil$, while $\F_{\Phi(t)}$ evolves in time in a non trivial way.

Of course, the results deduced here are strongly related to our choice of $H=H_0$. In particular we see that, even in the absence of interactions, a free dynamics for incompatible questions of the kind considered here produces a non trivial time evolution for the system. It is surely interesting to see what happens if $\lambda\neq0$, so that $H=H_0+\lambda H_I$, and try to understand which expressions for $H_I$ are reasonable in our context. We hope to be able to give some answer soon.

\section{Conclusions}\label{sectconclusions}

We have proposed a simple approach to the analysis of compatible and incompatible questions, based on the role of commutators for Hermitian operators and on the related Heisenberg-Robinson inequality. The approach proposed here simplifies significantly the one considered in \cite{bbkp}, and can be efficiently adopted when the dynamics of the system is considered. We have computed, for a simple but not trivial choice of the Hamiltonian driving the time evolution, how the mean values of the operators associated to the incompatible questions evolve, and we have considered, in many details, a simple example of the procedure.

Of course, this paper is just a first step towards a more detailed analysis of the dynamics in Decision Making processes: more general questions, not necessarily binary, and more general Hamiltonians, should be considered, together with more applications to concrete cases. These are only part of our future plans.

\section*{Acknowledgements}

The author acknowledges partial support from Palermo University and from G.N.F.M. The author really enjoyed several interactions with Prof. J. Busemeyer while writing this paper. Thanks!

\renewcommand{\theequation}{A.\arabic{equation}}

\section*{Appendix A:  A detailed analysis on the variance}

Because of its role in our analysis, we review briefly some useful facts on the expectation value and on the variance of an Hermitian operator $\Xh$ living in $\Hil=\mathbb C^2$ on a normalized vector $\Phi\in\Hil$. We call $x_1$ and $x_2$ the eigenvalues of $\Xh$, assuming that $x_1<x_2$, and $\varphi_1$ and $\varphi_2$ the corresponding eigenvectors: $\Xh \varphi_j=x_j \varphi_j$, $j=1,2$. As usual, we put 
$$
X_\Phi=\left<\Phi,\Xh\Phi\right>, \qquad (\Delta X_\Phi)^2=\left<\Phi,(\Xh-X_\Phi)^2\Phi\right>,
$$
where $\Phi=c_1\varphi_1+c_2\varphi_2$, $|c_1|^2+|c_2|^2=1$. Easy computations show that we can write
\be X_\Phi=|c_1|^2(x_1-x_2)+x_2,\qquad \Delta X_\Phi=(x_1-x_2)|c_1|\sqrt{1-|c_1|^2}.
\label{a1}\en
We see that $\Delta X_\Phi=0$ if $|c_1|=0,1$, which correspond to  $|c_2|=1,0$. Moreover, if $|c_1|=0$ then $X_\Phi=x_2$, while $X_\Phi=x_1$ if $|c_2|=0$. It is possible to prove the following inequalities, true for all $\Phi$:
\be
x_2\leq X_\Phi\leq x_1, \qquad 0\leq \Delta X_\Phi\leq \frac{x_1-x_2}{2}=:\Delta X_{max},
\label{a2}
\en
which show that $x_2+\Delta X_{max}<x_1$. Notice that $\Delta X_\Phi$ is not necessarily small, when compared to $x_1$ or $x_2$. Its magnitude, in fact, can easily increase when $x_1$ and $x_2$ are different enough, so that the difference $x_1-x_2$ is large. On the other hand, $\Delta X_\Phi$ is surely small if $x_1\simeq x_2$, which is not the case for the situation considered in Section \ref{sect2}, where $x_1=-x_2=1$.

\renewcommand{\theequation}{B.\arabic{equation}}

\section*{Appendix B:  Saturation of the Uncertainty relation}\label{AppB}

The starting point is the inequality
$$\Delta \F_\Psi\,\Delta \EE_\Psi\geq\frac{\left|\left<\Psi,[\Fh,\Eh]\Psi\right>\right|}{2},$$
in Section \ref{sectIQ}. It is interesting to discuss the possibility of finding a non trivial vector $\Phi$ saturating it, i.e. producing the equality
\be\Delta \F_\Phi\,\Delta \EE_\Phi=\frac{\left|\left<\Phi,[\Fh,\Eh]\Phi\right>\right|}{2}.
\label{b1}\en
For simplicity we restrict to the operator $\Eh$ in (\ref{add1}) with the particular choice $a=\cos\theta$, $b=\sin\theta$ and $\varphi=0$. Hence
$$
\Fh=\left(
\begin{array}{cc}
1 & 0 \\
0 & -1 \\
\end{array}
\right), \qquad \Eh=\left(
\begin{array}{cc}
\cos2\theta & \sin2\theta \\
\sin2\theta & -\cos2\theta \\
\end{array}
\right), \qquad [\Fh,\Eh]=2\sin2\theta\left(\begin{array}{cc}
0 & 1 \\
-1 & 0 \\
\end{array}
\right).
$$ 
The unknown state $\Phi$ is the following normalized column vector
$$
\Phi=\left(
\begin{array}{c}
\varphi_1 e^{i\theta_1} \\
\varphi_2 e^{i\theta_2} \\
\end{array}
\right),
$$
where $\varphi_j,\theta_j\in \mathbb{R}$, and  $\varphi_1^2+\varphi_2^2=1$. Straightforward computations show that
$$
\F_{\Phi}=\varphi_1^2-\varphi_2^2, \qquad \EE_\Phi=\cos2\theta(\varphi_1^2-\varphi_2^2)+2\sin2\theta\varphi_1\varphi_2\cos(\theta_1-\theta_2),
$$
while $(\Delta \F_\Phi)^2=1-\F_{\Phi}^2$ and $(\Delta \EE_\Phi)^2=1-\EE_{\Phi}^2$. Also, 
$$
\frac{1}{2}\left|\left<\Phi,[\Fh,\Eh]\Phi\right>\right|=2|\sin2\theta\varphi_1\varphi_2\sin(\theta_2-\theta_1)|.
$$
With these results, it is easy to see that (\ref{b1}) is satisfied if $\varphi_1=0$ or if $\varphi_2=0$. But these solutions correspond to $\Phi$ being an eigenstate of $\Fh$, which is obvious. Another solution which can be easily found is $\theta_1=\theta_2$, $\varphi_1=\cos\theta$ and $\varphi_2=\sin\theta$. But, with this choice, $\Phi$ is an eigenstate of $\Eh$, obvious again. We are interested in finding solutions which are not of this type. Rather than looking for the most general solution, we restrict here to the special case of $\theta=\frac{\pi}{4}$. Other choices can be considered similarly. Equation (\ref{b1}) becomes
$$
\cos^2(\theta_1-\theta_2)=4\varphi_1^2\varphi_2^2\cos^2(\theta_1-\theta_2),
$$
which admits two different type of solutions: (i) if $\cos(\theta_1-\theta_2)=0$ all choices of $\varphi_j$ are possible, if $\varphi_1^2+\varphi_2^2=1$. On the other hand, if  $\cos(\theta_1-\theta_2)\neq0$, then the only possible solutions are those with $\varphi_1^2=\varphi_2^2=\frac{1}{2}$. However, in this case, it is important to restrict to those values of $\theta_1$ and $\theta_2$ for which $\Phi$ is not an eigenstate of $E$, not to make the situation trivial.

Solutions of this kind do exist: for instance, if $\theta_1=\frac{\pi}{2}$, $\theta_2=0$, $\varphi_1=\frac{1}{2}$ and  $\varphi_2=\frac{\sqrt{3}}{2}$, we compute $(\Delta \F_\Phi)^2=\frac{3}{4}$, $(\Delta \EE_\Phi)^2=1$ and $\frac{1}{4}\left|\left<\Phi,[\Fh,\Eh]\Phi\right>\right|^2=\frac{3}{4}$. Hence $\Phi$ solves (\ref{b1}), and $\Phi$ is  an eigenstate of neither $\Eh$, nor $\Fh$. 

Another possible choice is the following:  $\theta_1=\frac{\pi}{4}$, $\theta_2=0$, $\varphi_1=\varphi_2=\frac{1}{\sqrt{2}}$. Then $(\Delta \F_\Phi)^2=1$, $(\Delta \EE_\Phi)^2=\frac{1}{2}$ and $\frac{1}{4}\left|\left<\Phi,[\Fh,\Eh]\Phi\right>\right|^2=\frac{1}{2}$. Again, we saturate (in a non trivial way),
the uncertainty relation for $\Eh$ and $\Fh$. This means that many {\em optimal states} do exist, at least in our particular case, and, we expect, also in more realistic systems relevant in Decision Making. In Quantum Mechanics, states which saturate the Heisenberg-Robinson inequality are rather important: coherent states, for instance, have this properties, \cite{gazeaubook}. We wonder if the states found in this section (or others, with similar properties) have some relevance in Decision Making. This is an open problem.

\end{document}